\documentclass[12pt]{article}
\usepackage[top=1in, bottom=1in, left=1in, right=1in]{geometry}
\usepackage[english]{babel}
\usepackage{atbegshi,cite}
\usepackage{amsmath,amssymb,amsbsy,amstext, amsthm, simplewick}
\usepackage{graphicx}
\usepackage{amsfonts}
\usepackage[small]{caption}
\usepackage{upgreek}
\usepackage[titletoc]{appendix}
\usepackage{setspace}
\usepackage{comment}
\usepackage[usenames,dvipsnames,table]{xcolor}
\usepackage[colorlinks=true,urlcolor=blue
,anchorcolor=blue,citecolor=blue,filecolor=blue,linkcolor=blue,menucolor=blue
]{hyperref}

\newcommand{\newc}{\newcommand}
\newc{\fpi}{f_{\pi}}
\newc{\etap}{\eta^{\prime}}
\newc{\llll}{\langle\lambda\lambda\rangle}
\newc{\FFd}{F^a\tilde F^a}
\newc{\qbar}{{\overline q}}
\newc{\TR}{{\rm Tr}}
\newc{\Kahler}{K\"ahler }
\newc{\Zbb}{{\mathbb Z}}
\newc{\Rt}{{{\mathbb R}^3}}
\newc{\Rf}{{{\mathbb R}^4}}
\newc{\Sth}{{{\mathbb S}^3}}
\newc{\SthSo}{{{\mathbb S}^3\times{\mathbb S}^1}}
\newc{\Stw}{{{\mathbb S}^2}}
\newc{\StwSo}{{{\mathbb S}^2\times{\mathbb S}^1}}
\newc{\So}{{{\mathbb S}^1}}
\newc{\zt}{{{\mathbb Z}_2}}
\newc{\RtSo}{{{\mathbb R}^3\times{\mathbb S}^1}}
\newc{\RfSo}{{{\mathbb R}^4\times{\mathbb S}^1}}
\newc{\scriminus}{{\cal I}^-}
\newc{\scriplus}{{\cal I}^+}
\newc{\mpl}{M_p}
\newc{\Ricci}{\mathcal{R}}
\newc{\bv}{\phi}
\newc{\calU}{{\cal U}}
\newc{\calK}{K}
\newc{\calUi}{{\cal U}^{-1}}
\newc{\calG}{{\cal G}}
\newc{\calI}{{\cal I}}
\newc{\calO}{{\cal O}}
\newc{\calQ}{{\cal Q}}
\newc{\calOb}{{\cal O}^\dagger}
\newc{\hphi}{{\hat\phi}}
\newc{\GeV}{\mathrm{GeV}}

\theoremstyle{plain}
\theoremstyle{plain} 
\theoremstyle{plain} 
\theoremstyle{plain}
\theoremstyle{plain}
\theoremstyle{plain}

\renewcommand{\title}[1]{{\Large\bf\flushleft{#1}}\vspace*{3ex}\\}
\renewcommand{\author}[2]{{\noindent\hspace*{2.5em}\large#1}
                     \footnote{Electronic mail: $\mathtt{#2}$}\\}

\newcommand{\beq}{\begin{equation}}
\newcommand{\eeq}{\end{equation}}

\newcommand\snowmass{\begin{center}\rule[-0.2in]{\hsize}{0.01in}\\\rule{\hsize}{0.01in}\\
\vskip 0.1in Submitted to the  Proceedings of the US Community Study\\ 
on the Future of Particle Physics (Snowmass 2021)\\ 
\rule{\hsize}{0.01in}\\\rule[+0.2in]{\hsize}{0.01in} \end{center}}
\begin{document}
\begin{titlepage}
\begin{flushright}
{\large 
~\\
}
\end{flushright}

\vskip 0.2cm

\begin{center}

{\large \bf Snowmass White Paper:\\
Strong CP Beyond Axion Direct Detection}

\vskip 1.4cm

{Nikita Blinov$^{(a)}$,
Nathaniel Craig$^{(b)}$,
Matthew J. Dolan$^{(c)}$,\\
Jordy de Vries$^{(d,e)}$
Patrick Draper$^{(f)}$,
Isabel Garcia Garcia$^{(g)}$,\\
Benjamin Lillard$^{(f)}$, 
Jessie Shelton$^{(f)}$}
\\
\vskip 1cm
{$^{(a)}$ Department of Physics and Astronomy, University of Victoria, Victoria, BC V8P 5C2, Canada}\\
{$^{(b)}$ Department of Physics, University of California, Santa Barbara, CA 93106, USA}\\
{$^{(c)}$ ARC Centre of Excellence for Dark Matter Particle Physics, School of Physics, University of Melbourne, Victoria 3010, Australia}\\
{$^{(d)}$ Institute for Theoretical Physics Amsterdam and Delta Institute for Theoretical Physics, University of Amsterdam, Science Park 904, 1098 XH Amsterdam, The Netherlands}\\
{$^{(e)}$ Nikhef, Theory Group, Science Park 105, 1098 XG, Amsterdam, The Netherlands}\\
{$^{(f)}$ Department of Physics and Illinois Center for the Advanced Study of the Universe, University of Illinois, Urbana, IL 61801, USA}\\
{$^{(g)}$ Kavli Institute for Theoretical Physics, University of California, Santa Barbara, CA~93106, USA}\\
\vspace{0.3cm}
\vskip 4pt

\vskip 1.5cm

\begin{abstract}
We sketch recent progress and promising future directions for research connected with the strong CP problem. Topics surveyed include axion dark matter substructure and its gravitational detection; axion model building and the quality problem; experimental tests of ultraviolet solutions; and connections to lattice QCD.

\end{abstract}

\end{center}
\snowmass

\vskip 1.0 cm

\end{titlepage}
\setcounter{footnote}{0} 
\setcounter{page}{1}
\setcounter{section}{0} \setcounter{subsection}{0}
\setcounter{subsubsection}{0}
\setcounter{figure}{0}



{\noindent \bf Introduction.}~~~The severity of what is known as the ``strong CP problem" has been increasing by about a factor of ten per decade since the  1950s.  In the  quantum chromodynamics (QCD)  Lagrangian governing the strong interactions, a phase parameter $\theta$ controls the violation of  the discrete symmetries of parity (P) and charge-parity (CP), and it feeds directly into  the 
electric dipole moment (EDM) observables of hadronic systems. Historically, Smith, Purcell, and Ramsey already obtained a bound equivalent to $|\theta|\lesssim 10^{-4}$ from measurements of the neutron EDM in the early 1950s, although they did not publish the results for several years~\cite{Smith:1957ht}. At that time it would still be two decades before QCD was understood to describe the strong interactions, and a few years more before $\theta$ was realized to be one of its physical parameters. Today, experimental limits on EDMs currently bound $|\theta| \lesssim  10^{-10}$~\cite{Abel:2020pzs}.
The strong CP problem is to find a reason why this phase is so small. It is a problem of technical naturalness in the sense of 't Hooft: the only putative Standard Model symmetries that $\theta$ breaks are P and CP, but these are not symmetries of the rest of nature, either. P and CP are both broken by the weak interactions, and further sources of CP violation beyond the Standard Model are needed to explain why the universe has more matter than antimatter. In contrast, for no apparent reason, the strong interactions preserve P and CP to fantastic precision.
 Lacking a Standard Model symmetry to explain this empirical fact, we are compelled to seek other dynamical solutions to the strong CP problem beyond the Standard Model.

At present there is no conclusive experimental evidence for any of the proposed solutions and considerable theoretical evidence against generic solutions: axion models need either a shockingly good global symmetry, a delicate arrangement of the initial conditions, a nonstandard cosmology, or other feature going beyond either vanilla effective field theory or simple string theoretic realizations of the Peccei-Quinn mechanism~\cite{Peccei:1977hh,Peccei:1977ur}; UV solutions like the Nelson-Barr mechanism~\cite{nelsoncp,barrcp,barrcp2} or left-right symmetric models~\cite{Beg:1978mt,mohapatrasenjanovic,Georgi:1978xz,Babu:1989rb,barrsenjanovic} likewise require careful construction to avoid unwanted residual corrections to $\theta$ or the introduction of even worse fine-tuning problems than the strong CP problem presents in the first place. Due to these curious circumstances, the strong CP problem remains a fertile and rewarding ground for  model-building and an experimental challenge with the potential to tell us far more about UV physics than one might have guessed from the consideration of the neutron EDM alone.

In this Snowmass whitepaper we highlight important near-term opportunities  for theory and experiment to develop tools to study the strong CP problem. We are purposefully not discussing  the development of new direct detection strategies for axion dark matter, a rich and vibrant subject addressed by other  contributions to Snowmass; rather, we focus on some complementary lines of inquiry of particular interest to the authors. Even in this regard we will not aim to be comprehensive.

{\noindent \bf Axion dark matter substructure.}~~~For any axion direct detection search, null results place limits on the spatial distribution of axion DM and must be combined with independent measures of the small-scale structure of DM in order to unambiguously test the axion DM hypothesis. In other words, if the distribution of DM is too clumpy, direct
  detection searches could come up empty and lead to erroneously strong exclusion limits. 
Both the ``pre-inflationary/high-$f_a$" scenario, favored by the simplest string models~\cite{Svrcek:2006yi,Arvanitaki:2009fg}, and the ``post-inflationary/low-$f_a$" scenario, favored by the simplest relic density considerations~\cite{Abbott:1982af,Dine:1982ah,Preskill:1982cy}, can give rise to heterogeneous distributions on small scales. In the latter case, clumpiness is due to the stochastic nature of the phase transition. 
In the former case, the usual overclosure problem of high-scale axions may be diluted by a late-decaying saxion partner, generically present in string models~\cite{Banks:2002sd}, and the associated modification to the expansion rate enhances small scale structure~\cite{Nelson:2018via,Blinov:2019jqc}. 
The early-time formation of substructure in these scenarios has been recently studied semi-analytically~\cite{Hardy:2016mns,Fairbairn:2017sil,Enander:2017ogx,Nelson:2018via,Visinelli:2018wza,Blinov:2019jqc,Ellis:2020gtq,Barenboim:2021swl} and in $N$-body simulations~\cite{Vaquero:2018tib, Buschmann:2019icd,Eggemeier:2019khm, Xiao:2021nkb,Buschmann:2021sdq,OHare:2021zrq}. These studies suggest that the small-scale matter power spectrum is significantly enhanced compared to $\Lambda$CDM, leading to the early formation of axion minihalos. Most of the DM mass is expected to be in these objects at early times~\cite{Eggemeier:2019khm,Blinov:2019jqc}. 
  
It is not, however, obvious what the distribution of axion DM is at late times relevant to direct detection. The simple semi-analytic arguments as well as the current suite of simulations (which extend to about $z\sim 100$~\cite{Eggemeier:2019khm} or $z\sim 20$~\cite{Xiao:2021nkb}), are unable to definitively answer this question. 
Current simulations provide valuable insight into the early stages of structure formation, but the very small size of the first-forming minihalos ($\sim 10^{-12} M_{\odot}$) means that $N$-body simulations that have fine enough resolution to track the evolution of these objects are limited to  small volumes relative to the size of the Milky Way.  While simulations suggest that minihaloes have good survival probabilities as subhaloes within larger dark matter haloes, minihaloes are vulnerable to tidal disruption in encounters with baryonic objects.  Providing a sharp prediction of this disruption rate and subsequent dark matter distribution is a severe challenge to $N$-body simulations, given their finite resolution in modeling both dark and baryonic matter.
%
While strides have been made in tackling this issue~\cite{2017JETP..125..434D,Tinyakov:2015cgg,Delos:2019tsl,Kavanagh:2020gcy} by constructing analytic models of star-minihalo interactions, there is still no robust prediction of the late-time distribution of axions in the galaxy based on a  first-principles dynamical simulation over cosmological times. This is therefore a major outstanding question that needs to be addressed in the coming years  through high-resolution $N$-body simulations and further analytic modeling. 

Even  if the enhanced small-scale structure renders Earth-based direct axion  detection ineffective, future astrophysical techniques like pulsar timing~\cite{Siegel:2007fz,Baghram:2011is,Kashiyama:2018gsh,Clark:2015sha,Dror:2019twh,Ramani:2020hdo,Lee:2020wfn,Delos:2021rqs} and photometric monitoring of highly-magnified stars (``cluster caustic microlensing'') may be able to observe axion minihalos through their gravitational interactions  alone~\cite{Dai:2019lud,Arvanitaki:2019rax,Blinov:2021axd}. These observables are thus an important complement to axion direct detection.
The impact of tidal disruption on the late-time microhalo population remains a key question for both pulsar timing  and cluster caustic microlensing.  Pulsar timing arrays are sensitive to the distribution of microhaloes within the Milky Way halo, similar to direct detection. Cluster caustic microlensing observations, on the other hand, are dominated by the distribution of microhaloes within a galaxy cluster, where the typical stellar encounter rate is smaller and therefore the expected tidal disruption is less severe \cite{Blinov:2021axd}.

{\noindent \textbf{Solutions to Axion Quality Problem.}}~~~In a prototypical QCD axion model, the axion is the pseudo-Nambu--Goldstone boson of an approximate global $U(1)_\text{PQ}$ Peccei-Quinn symmetry~\cite{Peccei:1977hh,Peccei:1977ur}, which is spontaneously broken at a high scale $f_a$. The effects of nonperturbative QCD break $U(1)_\text{PQ}$ explicitly, generating a periodic potential for the axion that is minimized when the axion expectation value $\langle a \rangle \sim - f_a \theta$ precisely cancels against $\theta$, restoring P and CP in the QCD vacuum. 
For this mechanism to produce a sufficiently small effective value of $\theta(a)$, the Peccei-Quinn symmetry must be very nearly conserved, with no sources of explicit $U(1)_\text{PQ}$ violation other than QCD itself. This requirement may be in tension with general expectations about quantum gravity, however:  gravitational effects are not expected to respect global symmetries. 
If $U(1)_\text{PQ}$ is broken in other sectors of the theory apart from QCD, the minimum of the axion potential shifts away from the desired value of $\theta(a) = 0$. Parametrizing the PQ-violating  potential by $\delta V(a) \sim Q f_a^4 \cos(a/f_a)$, to recover $|\theta| < 10^{-10}$ the factor $Q$ must be extremely small, $Q \lesssim 10^{-62} (10^{12}\text{GeV} / f_a)^4$. This is the axion quality problem: the $10^{-10}$ fine tuning associated with $\theta$ has only been shifted into a different (and more severe) fine-tuning in the corrections to the axion potential.

Several solutions have been proposed that realize sufficiently small $Q$, and recently there have been developments both in field- and string-theoretic model building. In ``accidental'' axion models the PQ-charged fields are also charged under some locally conserved symmetries, so that all gauge-invariant PQ-charged operators are necessarily higher-dimensional, and $Q \sim (f_a / M_p)^n$ is suppressed by multiple powers of the Planck mass $M_p$. The local symmetry may be discrete~\cite{Chun:1992bn,Carpenter:2009zs,Harigaya:2013vja,Chen:2021haa} or continuous~\cite{Cheng:2001ys,Hill:2002kq,Fukuda:2017ylt,Lee:2018yak,Darme:2021cxx,Nakai:2021nyf}. In composite axion models~\cite{Randall:1992ut, DiLuzio:2017tjx, Lillard:2017cwx, Lillard:2018fdt,Lee:2021slp,Lillard:2021dxd} the axion is a baryon-like particle, composed of quarks charged under a strongly coupled non-Abelian gauge group. In these models the confining dynamics often instigate the spontaneous breaking of $U(1)_\text{PQ}$, so that the scale $f_a$ is generated dynamically. 

String theory provides a different route to addressing the quality problem~\cite{Svrcek:2006yi}. Axions are ubiquitous in string theory, and corrections to the potential are generally exponentially small, $Q \sim e^{-S}$. It is quite plausible that there is a stringy axion coupling to QCD for which the $S$'s are all sufficiently large; this is the case for a class of flux compactifications with a relatively large number ($N > 17$) of axions, for example~\cite{Demirtas:2021gsq}. 
It has also been argued recently that having an axion with the right properties may be tied to more fundamental issues like the ``no global symmetries" dictum of quantum gravity~\cite{Heidenreich:2020pkc}; this is an interesting new approach to the problem bearing further investigation.

In general, constructing solutions to the axion quality problem in new contexts is an active area of current research. Many of the particle models designed to address the quality problem introduce new states with Standard Model charges that may be more easily constrained than the axion itself. Direct or indirect searches for these charged particles would provide new tests of their associated axion models.

{\noindent \bf Testing UV solutions to strong CP.}~~~
One class of solutions to the strong CP problem is based on the observations that $\theta$ can violate certain discrete symmetries and the renormalization of $\theta$ in the Standard Model alone is minuscule~\cite{ellisgaillard,nelsoncp,barrcp,barrcp2,Beg:1978mt,mohapatrasenjanovic,Georgi:1978xz,Babu:1989rb,barrsenjanovic}. These solutions represent a wholly different mechanism from Peccei-Quinn and are associated with a different set of model-building and experimental challenges. As far as the strong CP problem is concerned, developing tests of both classes of models should be regarded as equally important.

One aspect of this is simply to construct robust models that realize UV solutions without shifting the fine-tuning to some other corner of the larger theory. This is a nontrivial exercise~\cite{Dine:2015jga,Albaid:2015axa} but a few supersymmetric and composite examples are known~\cite{hillerschmaltz,Vecchi:2014hpa}, and in recent years there have been model-building innovations in interesting new directions~\cite{hook_cp_violation,hook_p_violation,Carena:2019nnd,Hall:2018let,Dunsky:2019api}. The bound on $\theta$ is so strong that two-loop precision can be necessary to determine what, if any, parameter space of an otherwise attractive model is viable. For example, in the simplest model utilizing generalized parity symmetries to control $\theta$, two-loop contributions to $\theta$ confine the parameter space to a lower-dimensional manifold, and also suggest that $\theta$ is not too much smaller than the present bound~\cite{deVries:2021pzl}. UV solutions to strong CP often predict additional direct contributions to quark and lepton EDMs that are close to present bounds. While such contributions to quark EDMs are highly correlated to contributions to $\theta$, contributions to lepton EDMs provide an additional probe. Anticipated improvements in sensitivity to both hadronic and leptonic EDM observables in the coming decade will provide a compelling test of UV solutions.

 In light of current experimental constraints, the most promising 
 discrete symmetry-based solutions 
 to strong CP rely on a see-saw implementation of the masses of all the light SM fermions, as recently discussed in~\cite{Craig:2020bnv}. An irreducible signature of these constructions is the presence of $W'$ and $Z'$ resonances, with current direct searches at the LHC providing the most stringent test and setting a lower bound on the parity-breaking scale $v' \gtrsim 18$ TeV~\cite{ATLAS:2019erb,ATLAS:2019lsy}. A future high energy collider could probe parity-breaking scales well above a $100$ TeV, and would be a decisive test of these solutions. Additional collider signatures include a fermionic top partner at the scale $v'$, as well as deviations in Higgs boson properties characterized by a mixing angle $\sin \alpha \sim v/v'$. If $W'$ and $Z'$ resonances are discovered at the LHC, pointing to a low parity-breaking scale, these additional signatures could be accessible at future colliders featuring higher energy and/or higher precision.

  Beyond colliders, the expected breaking of parity by gravitational effects highlights additional avenues to probe these models. For example, explicit breaking of parity by dimension-5  Planck-suppressed operators gives a contribution to $\theta$ small enough to comply with current constraints, but large enough to be observable in near-future experimental measurements if the breaking is maximal. On the other hand, if the explicit breaking is small, the production and collapse of domain walls in the early universe (associated to spontaneous and explicit breaking of parity respectively) would lead to a stochastic background of gravitational waves potentially observable in current and future low-frequency gravitational wave observatories. A discovery on either of these fronts would provide circumstantial evidence in favor of parity-solutions complementing colliders searches~\cite{Craig:2020bnv}.

There are also ways to test UV solutions indirectly through low-energy precision measurements, if we are fortunate enough to observe hadronic EDMs in low energy experiments or signs of CP-violating SMEFT operators at high energy colliders. Let us focus on the case of hadronic EDMs. The key question is whether the results are consistent with a ``pure $\theta$" scenario, where $\theta$ is the only source of hadronic CP violation, or whether additional dimension 6 operators are indicated~\cite{deVries:2018mgf,deVries:2021sxz}.  Both the Peccei-Quinn mechanism and UV solutions are consistent with a small nonzero $\theta$ term (since the elimination of $\theta$ is always imperfect in these models), but on quite general EFT grounds UV solutions are inconsistent with additional observably-large sources of CP violation.  The fundamental sources of CP violation can be disentangled by cross-correlating multiple EDMs, and the connection to the strong CP problem provides another fundamental-physics motivation for a phenomenological program of measuring the EDMs of light nuclei, atoms, and radioactive molecules. Improvements in hadronic and nuclear theory are needed to maximize the experimental reach, including lattice and nuclear structure computations to map operators in the microscopic theory to observables~\cite{deVries:2018mgf,deVries:2021sxz}; we elaborate on lattice connections next.

{\noindent \bf Interplay with lattice QCD.}~~~In recent years there has been substantial progress in first-principles lattice QCD computations of the neutron EDM in terms of the $\theta$ term and higher-dimensional sources~\cite
{Shindler:2021bcx}.
Already before 2017 several groups reported nonzero values of the neutron EDM from $\theta$, but these results turned out to be contaminated by spurious EDM contributions from mixing with the CP-even anomalous magnetic moment \cite{Abramczyk:2017oxr}. After subtracting the spurious terms, no statistically significant result remained. Since then, several groups redid the computation with different strategies. Ref.~\cite{Dragos:2019oxn} used the gradient flow, several lattice spacings, and larger-than-physical pion masses to compute the neutron EDM from $\theta$ and then interpolated to the physical pion mass by making use of the fact that the neutron EDM vanishes in the chiral and continuum limit. The results indicate a nonzero signal two standard deviations away from zero. However, other groups performed the same computation and did not find statistically significant evidence for a nonzero neutron EDM \cite{Alexandrou:2020mds,Bhattacharya:2021lol}. Therefore a sharp resolution to the question of exactly how large the neutron EDM is for nonzero values of $\theta$ remains an open but achievable target for near-term study. On the other hand, lattice  computations of the nucleon EDMs arising from quark EDMs have already achieved success~\cite{Bhattacharya:2015esa}, while work is in progress towards the computation of EDMs arising from quark chromo-EDMs and the three-gluon Weinberg operator \cite{Cirigliano:2020msr,Kim:2021qae,Shindler:2021bcx,Bhattacharya:2022whc}. Put together, these computations will be of great value  disentangling the microscopic sources of CP violation from (hopefully) future non-zero EDM measurements.

There are also other interesting questions closely tied to the strong CP problem that lattice QCD is in a unique position to address, related to the role of small instanton effects close to the QCD scale. There has already been significant progress in the computation of the topological susceptibility at finite temperature, relevant for the misalignment contribution to the axion relic density~\cite{berkowitz,bonatib,borsanyia,sharma,borsanyib}. At zero temperature, the contribution of small instantons to the light quark masses and terms in the NLO chiral Lagrangian could also be resolved with a variety of different techniques~\cite{Dine:2014dga} and interesting work along these lines was recently explored in~\cite{Alexandrou:2020bkd}.
 These questions are also of broader importance because, regardless of the mechanism that addresses the strong CP problem, they probe basic properties of real QCD and the validity of the semiclassical approximation at distances just below the confinement scale.

\section*{Acknowledgements}
PD and BL are supported by US Department of Energy under grant number
DE-SC0015655. JS is supported by Department of Energy CAREER grant DE-SC0017840. NC is supported by the US Department of Energy under grant number DE-SC0011702. JdV acknowledges support from the Dutch Research Council (NWO) in the form of a VIDI grant. MJD is supported by the Australian Research Council under grants CE200100008 and FT180100324.

\bibliographystyle{JHEP}
\bibliography{snowmass_strongCP}
\end{document}